\documentstyle[12pt]{article}
\normalbaselines
\begin{document}

\baselineskip=24pt

\bibliographystyle{unsrt}
\vbox{\vspace{6mm}}

\begin{center}
{\Large Spectrum of light scattered from a ``deformed" \\
Bose--Einstein condensate}
\end{center}

\bigskip

\begin{center}
Stefano Mancini$^{\dag}$ and  
Vladimir I. Man'ko$^{\ddag}$ 
\end{center}

\bigskip

\begin{center}

${}^{\dag}${\it Dipartimento di Fisica and Unit\`a INFM,\\
Universit\`a di Milano,
Via Celoria 16, I-20133 Milano, Italy}

${}^{\ddag}${\it P.N. Lebedev Physical Institute,\\
Leninskii Prospekt 53, Moscow 117924, Russia}

\end{center}

\begin{center}
(Date: May 25, 1999) 
\end{center}

\bigskip 
\bigskip 
\bigskip

\begin{abstract}
The spectrum of light scattered from a Bose--Einstein condensate 
is studied in 
the limit of particle-number conservation. 
To this end, a description in terms of deformed
bosons is invoked and this leads  
to a deviation from the usual predict spectrum's shape as 
soon as the number of particles decreases.
\end{abstract}

\bigskip

PACS number(s): 

03.65.Fd (Algebraic methods),

03.75.Fi (Bose condensation),

42.50.Ct (Quantum statistical description of interaction of light 
and matter)

\newpage

The recent achievements of Bose--Einstein condensate (BEC) with a 
gas of atoms confined by a
magnetic trap~\cite{exp} has stimulated renewed interest in the 
question as to what signatures 
Bose--Einstein condensation imprints in the spectrum of light 
scattered from atoms in such 
a condensate \cite{spectra,java}.

As well known, to deal with the dynamics of BEC gas the Bogolubov 
approximation in quantum
many-body theory~\cite{bog} is an efficient approach, in which the 
creation and annihiliation
operators for condensated atoms are substituted by $c$-numbers. 
One shortcoming of this method is
that the total atomic particle-number may not be conserved after 
the approximation. Or a symmetry
may be broken. To remedy this default, Gardiner~\cite{gardiner} 
suggested a modified Bogolubov
approximation by introducing phonon operators which conserve the 
total atomic particle number $N$
and obey the bosonic commutation relation in the case of $N\to\infty$. 
In this sense, this phonon
operator approach gives an elegant infinite atomic particle-number 
approximation theory for BEC
taking into account the conservation of the total atomic number.

Along this line, the case of finite number of particle has been 
recentely investigated~\cite{china}, 
and the algebraic method of treating the effects of finite 
particle number in the atomic BEC has been
developed. It results a physical and natural realization of 
the quantum group theory~\cite{bied} in the BEC systems, 
whose possibility was already suggested 
in~\cite{scripta1}, 
thought in a different manner.

Here, we shall use the deformed algebra 
to study the response of a condensate with finite number 
of atoms to the laser light 
and focus our attention on steady-state excitation.

We consider a system of weakly interacting Bose gas in a trap
and a classical radiation field interacting with these
two-level atoms, where $b^{\dag}$,
$b$ denote the creation and annihiliation operators for the atoms 
in the excited state;
$a^{\dag}$, $a$, the creation and annihiliation operators for 
the atoms in the ground state. These operators 
satisfy the usual bosonic commutation relations.
The Hamiltonian of the model reads
\begin{equation}\label{H1}
H=\hbar\varpi b^{\dag} b+ \hbar \left[g(t)b^{\dag} a
+g^*(t)b a^{\dag}\right]\,,
\end{equation}
where $g(t)$ is a time-dependent coupling coefficient for the 
(classical) laser field coupled
to those two states with level difference $\hbar\varpi$. Usually, 
the time dependence of
$g(t)$ is given by $g\exp(-i\Omega t)$, with $\Omega$ being the 
frequency of the laser beam.

Note that, with the above Hamiltonian, the total atomic 
particle number 
$N=b^{\dag} b+a^{\dag} a$ is conserved.
In the thermodynamic limit $N\to\infty$, the Bogolubov 
approximation~\cite{bog} is usually applied,
in which the ladder operators $a^{\dag}$, $a$ of the ground state 
are replaced by a c-number
$\sqrt{N_c}$, where $N_c$ is the number of the initial 
condensated atoms.
As a result, Eq.~(\ref{H1}) becomes the Hamiltonian of
a forced harmonic oscillator. 

Moreover,
we have to consider the bath of photon modes, beside the 
classical driving field,
so that the total Hamiltonian will be \cite{java}     
\begin{eqnarray}\label{H2}
H&=&\hbar\varpi b^{\dag} b+\hbar\sqrt{N_c}\left[g(t) b^{\dag}
+h.c.\right]
\nonumber\\
&&+\sum_k\Omega_k c_k^{\dag}c_k
+\hbar\sqrt{N_c}\sum_k \xi(k)\left[b^{\dag}c_k+h.c.\right]\,,
\end{eqnarray}
where $c_k$ represent radiation modes (of frequency $\Omega_k$) 
which constitute the bath and $\xi(k)$ is the coupling 
coefficient pertaining to the internal atomic states.
 
Now, by eliminating the heat-bath variables, in view of the Markov 
approximation, in the case of Hamiltonian~(\ref{H2}),
it is possible~\cite{qnoise} to obtain a quantum stochastic 
differential equation that describes the dynamics of the $b$ 
mode in the Heisenberg picture 
\begin{equation}\label{eqnotdef}
{\partial_t}\, b(t)=-i\Delta b(t)-ig\sqrt{N_c}-\Gamma b(t)
+\sqrt{2\Gamma}\,b_{\rm in}(t)\,,
\end{equation}
where $\Delta=\varpi-\Omega$, and 
$\Gamma$ is the damping rate.
Roughly, the latter is given by 
$\Gamma=\gamma\sqrt{N_c}$ \cite{java}, 
where $\gamma$ is the one-atom linewidth \cite{qnoise}.
Finally, $b_{\rm in}(t)$ is the vacuum noise operator
\begin{eqnarray}\label{corr}
\langle b^{\dag}_{in}(t)b_{in}(t') \rangle&=&\langle 
b_{in}(t)b_{in}(t') \rangle=0\,,\nonumber\\
\langle b_{in}(t)b^{\dag}_{in}(t') \rangle&=&\delta(t-t')\,.
\end{eqnarray}
The solution of Eq.~(\ref{eqnotdef}) is well known~\cite{qnoise},
and in the steady-state regime it becomes   
\begin{equation}\label{betainfty}
\langle b(t)\rangle \equiv \beta
=\frac{-ig\sqrt{N_c}}{\Gamma+i\Delta}\,,
\end{equation}
\begin{equation}\label{solnotdef}
\delta b(\omega)=\frac{\sqrt{2\Gamma}}{\Gamma+i\Delta}
b_{in}(\omega)\,,
\end{equation}
where the semiclassical approximation 
$b(t)=\beta+\delta b(t)$ has been used.
In (\ref{solnotdef}), $\delta b(\omega)$ is the Fourier 
component of the operator $\delta b(t).$

The spectrum of the light scattered from 
the atoms is given by the correlation 
of the operators $b^{\dag}(t)$ and $b(t)$~\cite{java}. 
Hence, in the steady state, the spectrum of fluctuations
$\langle \delta b^{\dag}(\omega)\delta b(\omega') \rangle$ 
results zero everywhere,
by virtue of (\ref{solnotdef}) and (\ref{corr}).
This means that in the long time limit, 
only the equal time correlations survive.

Let us now come back to 
the Bogolubov approximation~\cite{bog}. 
It destroyes the symmetry of 
Hamiltonian~(\ref{H1}), 
i.e., the conservation of the total particle number is 
violated because  $[N,H]\ne 0$.
Then, to preserve the property of the initial model,
it is possible to 
determine the following phonon operators \cite{gardiner}
\begin{equation}\label{Bdef}
B =\frac{1}{\sqrt{N}}a^{\dag}b\,,\quad B^{\dag} 
=\frac{1}{\sqrt{N}}a b^{\dag}.
\end{equation}
These operators obey a deformed algebra \cite{china}.
In fact, a straightforward calculation leads to the following 
commutation relation
\begin{equation}\label{cr1}
\left[B,B^{\dag}\right]=1-2\eta b^{\dag} b\,,
\end{equation}
where we have introduced a small operator parameter 
$\eta=1/N$, which for sufficientely large number of atoms is considered
as $c$-number.
The algebra defined by Eq.(\ref{cr1}) belongs to the
$f$-deformed algebra \cite{naples}, where in general the
deformed operator is related to the undeformed one through
an operator valued function $f$ as
\begin{equation}
B=bf(b^{\dag}b)\,.
\end{equation}
In our particular case, we have
\begin{equation}
f(b^{\dag}b)=\sqrt{1-\eta(b^{\dag}b-1)}\,,
\end{equation}
and for small deformation we get
\begin{equation}\label{Bb}
B\approx b\left[1-\frac{\eta}{2}
\left(b^{\dag} b-1\right)\right]\,.
\end{equation}
With the above in mind, the total Hamiltonian (\ref{H2}) should be 
rewritten as
\begin{eqnarray}\label{H3}
H&=&\hbar\varpi b^{\dag} b+\hbar\sqrt{N}\left[g(t) B^{\dag}
+h.c.\right]
\nonumber\\
&&+\sum_k\Omega_k c_k^{\dag}c_k
+\hbar\sqrt{N}\sum_k \xi(k)\left[B^{\dag}c_k+h.c.\right]\,.
\end{eqnarray}
Since now $N$ is a conserved quantity, we can consider 
it as a $c$-number 
(we suppose that it coincide with $N_c$, 
i.e., all the atoms are initially in the condensate).
Essentially, Eq.~(\ref{H3}) describes the damped dynamics 
of a deformed oscillator. 
This is a rather cumbersome problem to deal with, as shown 
in \cite{scripta3}.
Here, we simplify the treatment with the following argumentations: 
in the second term of r.h.s. of Eq.(\ref{H3}), the nonlinear 
character of $B$ must be 
taken into account, since it is
evidenciate by the radiation-field amplitude $g$; instead, in the 
last term of r.h.s. of Eq.(\ref{H3}), such
nonlinear character can be neglected due to the weak-coupling 
assumption with the heat bath.
Hence, the resulting effective Hamiltonian, in a frame rotating 
with the laser frequency, is 
\begin{eqnarray}\label{Heff}
H_{eff}&=&\hbar\Delta b^{\dag} b
+\hbar\sqrt{N}g\left( b^{\dag}+b\right)
-\hbar\frac{\sqrt{N}g\eta}{2}\left( b^{\dag}b^2+b^{\dag\,2}b\right)
\nonumber\\
&&+\sum_k\Omega_k c_k^{\dag}c_k
+\hbar\sqrt{N}\sum_k \xi(k)\left[b^{\dag}c_k+b c_k^{\dag}\right]\,,
\end{eqnarray}
where we have expressed $B$ in terms of $b$ by means of 
Eq.(\ref{Bb}).

The nonlinear quantum stochastic differential equation \cite{qnoise} 
describing 
the dynamics of the $b$-mode 
is now derived from Eq.~(\ref{Heff})
\begin{equation}
{\partial_t}\, b(t)=i\frac{\sqrt{N}g\eta}{2}\left(
b^2(t)+2b^{\dag}(t) b(t)\right)
-i\Delta b(t)-ig\sqrt{N}-\Gamma b(t)
+\sqrt{2\Gamma}\,b_{\rm in}(t)\,.
\end{equation}
It obviously reduces to linear equation~(\ref{eqnotdef}) as soon as 
$\eta\to 0$.

The steady state value of the field is given by the solution 
of the following equation
\begin{equation}
0=i\frac{\sqrt{N}g\eta}{2}\left(
\beta^2+2|\beta|^2\right)
-i\Delta \beta-ig\sqrt{N}-\Gamma \beta\,.
\end{equation}
Of course, the solution of the above equation will be 
different from that of 
Eq.~(\ref{betainfty}) (we refer to the latter as 
$\beta_{\infty}$), but
they approach each other as soon as $N$ increases, 
as can be seen in Fig.~1.

The dynamics of the small fluctuations is given by
\begin{equation}
{\partial_t}\, \delta b(t)={\cal A}\,\delta b(t)+{\cal B}\,
\delta b^{\dag}(t)
+\sqrt{2\Gamma}\,b_{\rm in}(t)\,,
\end{equation}
where
\begin{eqnarray}
{\cal A}&=&-i\Delta-\Gamma+i\sqrt{N}g\eta\left(
\beta+\beta^*\right)\,,\\
{\cal B}&=&i\sqrt{N}g\eta\beta\,.
\end{eqnarray}
In this case, the solution takes the form
\begin{equation}\label{deltab}
\delta\,b(\omega)=\frac{1}{\Xi(\omega)}\left\{
\left[i\omega-{\cal A}^*\right]\,b_{in}(\omega)
+{\cal B}\,b_{in}^{\dag}(\omega)\right\}\,,
\end{equation}
where
\begin{equation}\label{Xi}
\Xi(\omega)=|{\cal A}|^2-|{\cal B}|^2-\omega^2-i\omega 
({\cal A}+{\cal A}^*)\,.
\end{equation}

Finally, the spectrum, by means of Eqs.~(\ref{deltab}), 
(\ref{Xi}) and (\ref{corr}), reads
\begin{equation}\label{spec}
S(\omega)=\int d\omega'\;\langle \delta\,b^{\dag}(\omega)
\delta\,b(\omega')\rangle=\frac{|{\cal B}|^2}{|\Xi(\omega)|^2}\,.
\end{equation}
It is shown in Fig.~2 as a function of $N$ (besides $\omega$).
It trivially vanishes for $\eta\to 0$ (i.e., $N\to\infty$), 
otherwise it gives a 
signature of finite number of particles. 
More precisely, we may see that it shows a central peak 
(typical of the Lorentian shape) by decreasing the number 
of particles.
On the other hand, 
it would be interesting to study the transition from 
this structure 
to the characteristic Mollow triplet \cite{mollow} of a single 
trapped atom.
Unfortunately, our approximations are no 
longer valid for very 
few atoms, and one should devise a technique to solve 
completely the 
problem or investigate it numerically.
This is a plan for the future study.

Summarizing, we have seen that the particle-number 
conservation in BEC 
requires a deformation of the bosonic field, hence the 
introduction of nonlinearity
\cite{naples}, which may lead (in the limiting case of small
number of
particles) to observable effects on a probe light field.
Beyond the oversimplified model used, we retain the 
measurement of the light spectrum in presence of few 
condensed atoms a promising experimental challenge.
On the other hand, the use of a BEC with small number of 
atoms would be the 
subject of next generation experiments~\cite{tino}. 

The same aim could be pursued in  
elementary particle field as well.
In fact, the BEC may also describe
the final state of pions in high-energy-heavy-ion 
collisions~\cite{pion1,pion2}.

\section*{Acknowledgements}
V. I. Man'ko is grateful to Russsian Foundation for Basic Research under the 
Project No. 99-2-17753.

\newpage

\baselineskip=24pt

FIGURE CAPTIONS

Fig.~1. A plot of the quantity $||\beta|-|\beta_{\infty}||$ 
as a function of $N$.
Values of the parameters are: $\Delta=0$ and $g=2.5\,\gamma$.
Furthermore, $\arg[\beta]=\arg[\beta_{\infty}]=\pi/2$ $\forall N$.

Fig.~2. The spectrum $S$ as a function of $\omega$ and $N$.
Values of parameters as in Fig.~1.


\begin{thebibliography}{99}


\bibitem{exp} 
M. H. Anderson J. R. Ensher, 
M. R. Matthews, C. E. Wieneman, E. A. Cornell, Science
{\bf 269}, 198 (1995);
K. B. Davies, M. -O. Mewes, M. R. Andrews. 
N. J. van Druten, D. S. Durfee, D. M.
Kurn, W. Ketterle,  Phys. Rev. Lett. {\bf 75}, 3969 (1995);
C. C. Bradley, C. A. Sackett, J. J. Tollett and 
R. G. Hulet, Phys. Rev. Lett. {\bf 75},
1687 (1995).

\bibitem{spectra}
M. Levenstein and L. You,  Phys. Rev. Lett. {\bf 71}, 1339 (1993);
L. You, M. Lewenstein and J. Cooper, 
Phys. Rev. A {\bf 50}, R3565 (1994);
R. Graham and D. F. Walls, Phys. Rev. Lett. {\bf 76}, 1774 (1996).

\bibitem{java}
J. Javanainen, Phys. Rev. Lett.
{\bf 72}, 2375 (1994); 
{\bf 75}, 1927 (1995).

\bibitem{bog}
N. N. Bogolubov, J. Phys. (Moscow) {\bf 2}, 23 (1947).

\bibitem{gardiner}
C. W. Gardiner, Phys. Rev. A {\bf 56}, 1414 (1997).

\bibitem{china}
C. P. Sun, S. Yu and Y. B. Gao, eprint qunt-ph/9809079. 

\bibitem{bied}
L. C. Biedenharn, J. Phys. A {\bf 22}, L873 (1989);
A. J. Macfarlane, J. Phys. A {\bf 22}, 4581 (1989).

\bibitem{scripta1}
S. Mancini, V. I. Man'ko and P. Tombesi,
Physica Scripta {\bf 57}, 486 (1998).

\bibitem{qnoise}
C. W. Gardiner, {\it Quantum Noise} (Springer, Heidelberg, 1991).

\bibitem{naples}
V. I. Man'ko, G. Marmo, F. Zaccaria and E. C. G. Sudarshan,
Physica Scripta {\bf 55}, 528 (1997).

\bibitem{scripta3}
S. Mancini,
Physica Scripta {\bf 59}, 195 (1999).

\bibitem{mollow}
B. R. Mollow, Phys. Rev. {\bf 188}, 1969 (1969).

\bibitem{tino}
G. M. Tino, private communication.

\bibitem{pion1}
Proceedings of the Quark Matter'96 Conference, 
(P. Braun-Munzinger {\it et al.}, eds.),
Nucl. Phys. {\bf A610}, 1c-565c (1996);
Proceedings of the Strangeness in 
Hadronic Matter'96 Conference, (T. Csorgo {\it et al.,} eds.), 
Heavy Ion Physics {\bf 4}, 1--440 (1996).

\bibitem{pion2}
T. Cs\"orgo and J. Zimanyi, 
eprint hep-ph/9705432,
eprint hep-ph/9705433.



\end{thebibliography}
\end{document}